# Calculations of Energy Losses due to Atomic Processes in Tokamaks with Applications to the ITER Divertor

D. Post[1], J. Abdallah[2], R. E. H. Clark[2], and N. Putvinskaya[1]

[1]*ITER Joint Central Team, San Diego, CA,*

[2]*Los Alamos National Laboratory, Los Alamos, NM*

**Abstract**

Reduction of the peak heat loads on the plasma facing components is essential for the success of the next generation of high fusion power tokamaks such as the International Thermonuclear Experimental Reactor (ITER)[1]. Many present concepts for accomplishing this involve the use of atomic processes to transfer the heat from the plasma to the main chamber and divertor chamber walls and much of the experimental and theoretical physics research in the fusion program is directed toward this issue. The results of these experiments and calculations are the result of a complex interplay of many processes. In order to identify the key features of these experiments and calculations and the relative role of the primary atomic processes, simple quasi-analytic models and the latest atomic physics rate coefficients and cross sections have been used to assess the relative roles of central radiation losses through bremsstrahlung, impurity radiation losses from the plasma edge, charge exchange and hydrogen radiation losses from the scrape-off layer and divertor plasma and impurity radiation losses from the divertor plasma. This anaysis indicates that bremsstrahlung from the plasma center and impurity radiation from the plasma edge and divertor plasma can each play a significant role in reducing the power to the divertor plates, and identifies many of the factors which determine the relative role of each process. For instance, for radiation losses in the divertor to be large enough to radiate the power in the divertor for high power experiments, a neutral fraction of $10^{-3}$ to $10^{-2}$ and an impurity recycling rate of $n_e\tau_{recycle}$ of $\sim 10^{16}$ s m$^{-3}$ will be required in the divertor.

# 1. Introduction

If all of the heating power in large fusion experiments such as the International Thermonuclear Experimental Reactor (ITER)[2] strikes the divertor plates, the peak heat loads on plasma facing components will be very high. These loads can be characterised in terms of the heat flux perpendicular to the open flux tubes at the plasma edge, $Q_{\parallel} \approx \dfrac{P_{heat}}{2\lambda 2\pi a} \dfrac{q_{\psi}}{\sqrt{\dfrac{1+\kappa^2}{2}}}$, where $q_{\psi}$ is the MHD safety factor, $\lambda$ is the radial decay length of the power at the midplane, a is the plasma minor radius, and $\kappa$ is the plasma elongation. For typical ITER parameters of 300 to 600 MW of alpha heating, $Q_{\parallel} \approx 2000$—4000 MW/m$^2$. The peak heat flux on a divertor plate can be reduced by expansion of the flux surfaces in the vicinity of the X point and by the tilt of the field lines $\dfrac{B_p}{B_T}$ and inclination of the divertor plate to 20-40 MW/m$^2$. Such heat loads are too high to allow the development of a sound divertor design because the surface temperatures are too high for the thicknesses required for the divertor be able to survive transient events such as plasma disruptions. The "wetted area" of the divertor plates, $2\pi R \lambda \times 2$ (two divertor legs) $\times 4$ (flux surface expansion factor) $\approx 4$ m$^2$ with a radial decay length, $\lambda$, for the power of 0.01 cm, is much smaller than the total surface area available in the divertor chamber (200—400 m$^2$) or in the main plasma chamber (1200 m$^2$). The peak heat loads can be reduced to the 0.6—4 MW/m$^2$ range if the power can be spread out on the walls of the divertor chamber or radiated from the main plasma. The ITER divertor is designed to maximise the role of atomic processes of charge exchange, hydrogen and impurity line radiation, ionisation, and elastic collisions between the recycling gas and the plasma in the diverted plasma to spread out the heat and momentum[2] (Figure 1).

Conditions where atomic processes have dispersed the heat and momentum have been realised on a number of tokamaks, including divertor experiments[3], and limiter experiments[4,5], but with lower power levels than needed for a next step experiment such as ITER. The challenge is to develop a divertor concept where these effects are strong enough to reduce the energy flux on the divertor plate by a factor of at least five and preferably to ten or more. There is an active program to develop and validate divertor simulations with the data from present experiments and use the simulations to analyse and assess divertor concepts for ITER[6]. The modelling results show some general trends. In particular, the calculational results indicate that charge exchange losses and hydrogen radiation losses are usually relatively small ($\leq 10$—20%), but that impurity radiation can be large if the conditions are optimal. The sophisticated computer models are very complex and time-consuming so that it is difficult to obtain a large number of parameter scans when impurity radiation is included. Similarly, the results of divertor experiments are complex and the experimental run time limited. To identify and evaluate the key issues involved in these experiments and calculations, we have assessed the potential role of the major candidate processes which might reduce the peak heat loads in the divertor: bremsstrahlung radiation from the central plasma, impurity radiation losses from the plasma edge, charge exchange and hydrogen radiation losses from the divertor plasma and impurity radiation from the divertor plasma. These processes all rely heavily on atomic collision effects, and we have assembled and assessed the most recent atomic data to use in this analysis. A solution of the power exhaust requirement for fusion experiments such as ITER will likely involve a mixture of all of these processes, with bremsstrahlung from the plasma center and impurity radiation from the plasma edge and the divertor playing the major roles.

The conditions under which impurity radiation in the divertor might be important is illustrated in Figure 2. This divertor concept, proposed for ITER, utilises baffles to confine the recycling neutral hydrogen and impurity gases in the divertor chamber and facilitate momentum exhaust. The louvers are semi-transparent to the neutrals and ensure that the recycling neutral flux is relatively uniform on the divertor plasma, and that the fast neutrals formed by charge exchange can transfer their momentum to the wall before striking other neutrals. The recycling neutrals are ionised in an ionisation front that stretches from the outer baffle to the central dome. The radiation zone would be located upstream of the recycling zone and the energy in the divertor is radiated onto water cooled louvers parallel to the plasma. The remaining plasma energy falls on dump targets at the bottom of the divertor chamber.

## 2. Bremsstrahlung Losses

Bremsstrahlung plays only a small role in present experiments due to high heating power densities that present experiments need to balance the transport losses. Experiments on Doublet III-D (DIII-D)[7] with 20 MW of heating has a power density of 20 MW / 23 m$^3$ ≈ 0.85 MW/m$^3$ compared to ITER with 300 MW / 2200 m$^3$ ≈ 0.13 MW/m$^3$. The electron temperatures and densities are larger and the local transport losses are smaller in ITER so that Bremsstrahlung can play a larger role than in present experiments.

The ratio of Bremsstrahlung losses, $P_{Brem} = C_B \, n_e^2 Z_{eff} T^{1/2}$, and alpha heating,

$$P_\alpha = \frac{n_e^2}{4} f_{DT}^2 \langle \sigma v \rangle_{DT} E_\alpha \; , \quad \frac{P_{Brem}}{P_\alpha} = \frac{4 C_B \, Z_{eff} T^{1/2}}{f_{DT}^2 \langle \sigma v \rangle_{DT} E_\alpha} \quad \text{scales like} \quad \frac{1}{T^{1.5}} \; since \; \langle \sigma v \rangle_{DT} \approx T^2 \; \text{and decreases}$$

with temperature. $P_{Brem}/P_\alpha$ is proportional to $\dfrac{Z_{eff}}{f_{DT}^2} \approx \dfrac{1 + 2 f_\alpha + \sum Z(Z-1) f_Z}{1 - 2 f_\alpha - \sum Z f_Z}$ where $f_\alpha = n_a/n_e$ and $f_{Be} = n_{Be}/n_e$, and increases with the impurity level and Z.

For sample operating conditions for ITER with $f_\alpha \approx 0.05$—0.15 and $f_{Be} \approx 0.1 \, f_\alpha$, $0.15 \le P_{Brem}/P_\alpha \le 0.6$ with a nominal value of 0.3 for $f_\alpha \approx 0.1$, $f_{Be} \approx 0.01$ (Figure 3). Bremsstrahlung losses play a minor role in present experiments due to the lower densities and temperatures and high heating power densities compared to fusion experiments such as ITER. For a 20 MW heating experiment ( ~ 0.8 MW/m$^3$) such as DIII-D with 1% carbon concentration, $P_{Brem}/P_{heat}$ ranges from 1 to 4 % . For ITER with 2% Be (no He) with 50 MW of auxiliary heating and no alpha heating, the bremsstrahlung losses would be very comparable to the auxiliary heating for temperatures above 7 or 8 keV.

## 3. Charge Exchange and Hydrogen Radiation Losses

Charge exchange is a potential mechanism to transfer the power from the divertor plasma to the side walls. Several factors, however, intervene to limit its effectiveness. Ionisation rates are comparable to charge exchange rates except for $T_e \leq 3$—$4$ eV (Figure 4). For densities of $\sim 10^{20}$ m$^{-3}$ (typical of most divertor plasmas), the recycling neutrals are not able to penetrate very far into the divertor plasma since, even in the limit where $\lambda_{ioniz} \ll \lambda_{cx}$,

$$\lambda_{eff} \approx \frac{1}{n}\sqrt{\frac{v_o v_o}{\langle\sigma v\rangle_{CX}\langle\sigma v\rangle_{ioniz}}} \approx \frac{0.0065}{n(10^{20}m^{-3})} cm \quad [8]$$

for $v_o \approx 10^4$ m/s and $\langle\sigma v\rangle \approx 1.5\ 10^{-14}$ m$^3$/s (including collisional radiative effects[9]) is small compared to the dimensions of the plasma. A simple model calculation comparing the heat flux lost via charge exchange and the heat flux on the plate indicates that charge exchange losses will be relatively small for realistic parameters from the divertor plasma between the X point and the divertor plate. The heat flux on the plate $Q_{plate}$ (MW/m) per unit toroidal length (Figure 5) is approximately $Q_{plate} = n_d v_d (T_d \gamma + 20 eV)\Delta \frac{B_{pol}}{B_{tor}}$ where $n_d$, $T_d$ and $v_d$ are the plasma density, temperature and velocity at the divertor plate, 20 eV is the sum of the hydrogen ionisation potential with some allowance for hydrogen radiation, $\Delta$ is the poloidal width of the divertor plasma and $B_{pol}$ and $B_{tor}$ are the poloidal and toroidal magnetic fields in the divertor. The charge exchange losses per unit toroidal length can be estimated as

$$Q_{CX} = \frac{n_o v_o}{4} 2 l \frac{3}{2} T_u r\, G \text{ where } G \equiv \min\left(\frac{2\lambda_{eff}}{\Delta}, 1\right),$$

where $n_o$ and $v_o$ are the neutral density and velocity, $\frac{3}{2}T_u$ is an estimate of the upstream average ion kinetic energy (which could be larger than $\frac{3}{2}T_u$ if the upstream flow speed is near sonic or super-sonic), $l$ is poloidal height of the divertor plasma and r is the plasma "reflectivity", the fraction of the neutrals incident on the plasma which come back due to charge exchange. Penetration effects are included by estimating the fraction of the divertor width which the neutrals can penetrate and weighting the losses by the factor G. The reflectivity can be estimated as $r \approx \frac{\langle\sigma v\rangle_{CX}}{\langle\sigma v\rangle_{CX} + \langle\sigma v\rangle_{ionization}}$. $n_o v_o$ and $n_d v_d$ are linked by particle flux balance: $\frac{n_o v_o}{4}(1-r)2l = n_d v_d \frac{B_p}{B_t}\Delta$ where (1-r) is the fraction of particles ionized and r is the fraction reflected by charge exchange collisions. Using particle flux balance, the fraction of energy lost to charge exchange is $f_{CX} = \frac{Q_{CX}}{Q_{CX} + Q_{plate}} = \frac{\frac{3}{2}T_u G r}{\frac{3}{2}T_u G r + (T_d \gamma + 20 eV)(1-r)}$. $\lambda_{eff}$ can be evaluated assuming pressure balance along the field lines: $n_s T_s = n_u T_u \Rightarrow \lambda_{eff} \approx \frac{T_u}{n_s T_s}\sqrt{\frac{v_o v_o}{\langle\sigma v\rangle_{CX}\langle\sigma v\rangle_{ioniz}}}$.

This analysis (Figure 4) indicates that the fraction of energy lost by charge exchange is about 15% or less for $T_u \leq 180$ eV for ITER conditions ($n_s = 10^{20}$ m$^{-3}$, $T_s = 260$ eV, $\Delta = 0.1$ m), and about 10 % for the most extreme conditions where $T_u \approx T_s$, with similar results for DIII-D ( $n_s = 5 \times 10^{19}$ m$^{-3}$, $T_s = 88$ eV, $\Delta = 0.1$ m). The chief factor reducing the charge exchange losses is the poor penetration of the neutrals to regions with high ion temperatures. Pressure balance leads to an increase in the density as the temperature falls toward the divertor (except near the divertor plate). The resulting increase in density increases the ionisation rate, and decreases the ionisation and charge exchange mean free paths.

Hydrogen radiation is also a potential energy loss from the scrape-off layer between the X point and the divertor plate. However, several effects limit the amount of hydrogen radiation. Unlike impurities, only recycling neutral hydrogen atoms from the wall will radiate and they must penetrate into the SOL plasma. At high densities ($\geq 10^{20}$ m$^{-3}$), hydrogen ionisation is enhanced and hydrogen radiation is suppressed due to multiple collisions with electrons[9]. The relative magnitude of hydrogen radiation losses due to the neutral flux on the sides of the divertor plasma (Fig. 2) can be estimated using our charge exchange loss model by replacing $3/2\, T_u$, the maximum energy lost per charge exchange event, by the radiated loss per ionisation.

$$f_{H-rad} = \frac{Q_{rad}}{Q_{rad} + Q_{plate}} = \frac{E_u G}{E_u G + T_d \gamma + 13.6\, eV}$$ . For the reference ITER and DIII-D cases ($n_s = 10^{20}$ m$^{-3}$, $T_s = 120$ and $260$ eV, and $\Delta \sim 0.1$ m, respectively), the lack of penetration and the lower emissivity at higher densities limit the hydrogen radiation to a few percent (Figure 6). The poor penetration of the neutral hydrogen and the suppression of the hydrogen radiation each play a role. An additional limit on the effectiveness of hydrogen radiation is imposed by the finite opacity of the neutral cloud for the dominant lines of hydrogen which reduces the flux of hydrogen radiation as well as increases the effective ionisation rate. $\lambda_{absorption}$ is $a/n_o(10^{20}$ m$^{-3})$ where a=0.002 m for $L_\alpha$ and .0004 m for $H_\alpha$. A simulation of model DIII-D conditions indicates that for densities in the $10^{20}$ m$^{-3}$ range, the flux of hydrogen radiation can be reduced by a factor of 2 or more[10].

## 4. Impurity Radiation from the main plasma edge

Impurity radiation from the main plasma and from the divertor plasma has the potential to spread out the heating power over the main chamber and divertor chamber walls and thereby reduce the peak heat loads. The temperature and density ranges for these impurities are in the 1 eV to 5000 eV and $10^{19}$ to $10^{22}$ m$^{-3}$ range. The progress in atomic physics during the last 15 years now allows relatively accurate calculation of impurity ionisation, recombination and excitation rate coefficients including direct and indirect ionisation and very detailed treatments of dielectronic recombination and excitation using collisional-radiative models which can treat meta-stable levels and include density effects[11]. Detailed calculations of the impurity emission rates for Be, B, C, Ne and Ar including collisional radiative effects have recently been carried out[12]. These new rates are more accurate for the nearly neutral species that exist at temperatures of several 100 eV than the rates that have been previously used before from the ADPAK code[13,14] (Figure 7). The ADPAK rates are based on energy levels derived from a screened hydrogenic model and employ scaled oscillator strengths and recombination rates so the differences are to be expected. The differences are even larger at low temperatures for higher Z elements such as Krypton. To assess the potential role of Krypton as an impurity feed gas, we have extended our previous calculations of the lower Z elements[12] to Krypton for $T_e$ from 1 to 200 eV. We find that the ADPAK rates for Krypton below 100 eV are up to 100 times larger than the more accurate calculations, but that the ADPAK rates above ~100 eV are reasonably accurate because Kr is then sufficiently ionized that the hydrogenic models used in ADPAK are adequate. We have therefore used our detailed calculations for Krypton up to 200 eV and the ADPAK rates for $T_e$ greater than 200 eV (Figure 8).

Many present experiments are able to radiate almost all of the heating power from the main plasma edge inside the last closed flux surface[4,5]. While edge radiation can be important in reducing the energy into the divertor, there are some potential drawbacks to exhausting all of the power by edge radiation. Large radiation losses from the main plasma will increase the heat flux on the first wall which is already near the engineering limits (~ 0.5 MW/m$^2$) for components that can only be replaced infrequently. If the edge density is not sufficiently high, the plasma volume required to radiate the power will be large, potentially requiring an increase in the minor radius. Large amounts of edge radiation may also adversely affect confinement by reducing the power across the separatrix below the threshold needed to reach the H-mode. Using the scalings developed from the ITER H-mode threshold database, $P_{H-threshold} = 0.025 n_e^{0.75} B_T S$ $or$ $0.4 n_e B_T R^{2.5}$ where $n_e$ is the line averaged density, $B_T$ is the toroidal field, S is the plasma surface area, and R is the major radius in units of MW, $10^{20}$ m$^{-3}$, T, m$^2$ and m[15], the power needed across the separatrix is 100 to 400 MW for ITER conditions(depending on the edge conditions), which implies that at least 100 MW will need to be exhausted by the divertor. Also, a cool edge may lead to low densities in the divertor, thereby increasing the pumping requirements for He exhaust.

The issues can be characterised using a simple model for radial energy transport at the plasma edge (r ≈ a)[16-18]. :

$$Q_\perp = \kappa \frac{\partial T}{\partial r} \quad \text{and} \quad \frac{\partial Q_\perp}{\partial r} = -n_e n_z L_z(T_e) \Rightarrow Q_\perp \frac{\partial Q_\perp}{\partial r} = -n_e^2 f_z L_z(T_e) \kappa \frac{\partial T}{\partial r}$$
$$\Rightarrow Q_\perp^2 \approx 2 \int_0^{T_c} n_e^2 f_z L_z(T_e) \kappa \, dT_e \quad (1)$$

To simplify the analysis and obtain a range for the achievable values of $Q_\perp$, we can assume that $n_e(r)$ is roughly constant near the edge, or at least has an average value $n_e$, and that $\kappa_\perp$ ~ constant which is roughly consistent with present experiments (e.g.[19]). This scaling is similar to the "INTOR" scaling since $n\chi_\perp \sim \kappa_\perp$ so that $\chi_\perp \approx n^{-1}$. This is roughly consistent with the observation that $\chi_\perp$ peaks at the plasma edge as $n_e$ drops. With H-mode operation in DIII-D in the scrape-off layer near the edge, $\chi_\perp \sim 0.25$—0.5 m$^2$/s [7], while for the L-mode or ELMy H-mode conditions, $\chi_\perp$ can be as large as 2 m$^2$/s(c.f.[20]). For edge densities between 5 and $10 \times 10^{20}$ m$^{-3}$,and $\chi_\perp$ between 0.25 and 2 , $\kappa_\perp \approx 0.125$—$2 \times 10^{20}$ m$^{-1}$ s$^{-1}$ . With these assumptions, the integral becomes: $Q_\perp^2 \approx 2 n_e^2 f_z \kappa \int_0^{T_c} L_z(T_e) \, dT_e$, where the upper limit on the temperature integral is the temperature on the inside boundary of the radiating layer, which for ITER would be in the 2 to 5 keV range. This scaling has $Q_\perp \approx n_e \sqrt{(\kappa f_z)}$, weaker than the volume loss rate with $P \approx n_e^2 f_z$. The integral has been evaluated for Be, B, C, Ne and Ar using a collisional radiative model, for Fe using ADPAK data and for Kr using a mixture of a detailed atomic model and ADPAK data (Figure 9)[12]. The integral ceases to increase with $T_e$ as each impurity becomes fully ionized, so that higher Z impurities such as Fe and Kr radiate more strongly than lower Z impurities. The advantage of the higher radiation rates, however, is off-set by the lower allowed fraction of the impurity.

The surface area of the ITER plasma is ~ 1200 m$^2$ so that $Q_\perp$ for ITER is 0.25—0.5 MW/m$^2$ for heating powers of 300 to 600 MW reaching the plasma edge. Assuming that the edge density is 5—10 × 10$^{19}$ m$^{-3}$ and that the impurity fraction is 1/3 of the "fatal fraction" for which the impurity radiation losses equal the alpha heating power[21], candidate low Z impurities radiate only about ~ 0.1 MW/m$^2$ (Table 1) and medium Z impurities up to ~ 0.3 MW/m$^2$ for sufficiently high edge densities and high edge conductivities.

Table 1  Radiation from the plasma edges ( $Q_\perp$ ~ 0.2 to 0.4 MW/m$^2$ required for ITER)

|  | Be | C | Ne | Ar | Fe | Kr |
|---|---|---|---|---|---|---|
| fatal fraction | 0.14 | 0.07 | 0.025 | 0.0054 | .0027 | .0017 |
| 1/3 fatal fraction | 0.05 | 0.023 | 0.008 | 0.002 | .0009 | .0006 |
| $Q_\perp/(n_e (\kappa_\perp f_z)^{0.5})$MWm* | 0.1 | 0.4 | 1.0 | 3.0 | 7 | 10 |
| $Q_\perp$ (MW/m$^2$) for $5 \times 10^{19} \le n_e < 10^{20}$ and $0.125 \le \kappa \le 2$ | 0.004 to 0.03 | 0.01 to 0.085 | 0.016 to 0.13 | 0.02 to 0.19 | 0.04 to 0.3 | 0.05 to 0.35 |

*$\kappa_\perp$ in 10$^{20}$ m$^{-1}$ s$^{-1}$ , $n_e$ in 10$^{20}$ m$^{-3}$

On TEXTOR, $Q_\perp$'s of 0.1 MW/m$^2$ have been radiated with stable condition and little or no confinement degradation[4]. Our analysis assumes that the impurities are in coronal equilibrium. Rapid recycling of impurities from the limiter and wall in TEXTOR plays a large role in enhancing the radiation over coronal equilibrium values. The ITER divertor is designed to localise the recycling of neutrals and impurities in the divertor chamber, so that one cannot be assured that the same level of enhancement will occur in ITER. Edge impurity radiation will thus play an important role in transferring some of the heating power to the first wall but, for the reasons outlined above, at least 100 MW or more will cross the separatrix to the scrape-off layer and divertor.

## 5. Impurity Radiation from the Divertor Plasma

The ITER divertor concept is based on the use of impurity radiation in the divertor chamber to transfer the power from the divertor plasma to the divertor chamber walls (Figure 2). Such "detached" operation has been produced on many experiments divertor experiments by a combination of gas puffing and injection of gaseous impurities such as Ne or Ar[3]. Detached conditions appear to be brought about by impurity radiation from the edge plasma which lowers the temperature and increases the density in a condensation instability to form a "MARFE-like" state, followed by momentum loss by charge exchange and elastic collisions which allow neutral atoms to transfer the plasma momentum to the walls(c.f. [22]). The plasma temperature must be reduced to ~ 5 eV for charge exchange and elastic collisions to be significant.

The conditions needed to radiate the required energy in the divertor can be assessed in a similar fashion to edge radiation by changing the form of the thermal conductivity to account for parallel heat conduction and using pressure balance along the field lines. The equations become:

$$\frac{\partial Q_\parallel}{\partial x} = -n_e n_z L_z(T_e) \quad \text{M} \quad Q_\parallel = -\kappa_o T_e^{2.5} \frac{\partial T_e}{\partial x} \quad \text{M} \quad p_e = n_e T_e \Rightarrow$$

$$\frac{1}{2} \frac{\partial Q_\parallel^2}{\partial x} \approx \frac{p_e^2}{T_e^2} \kappa_o f_z L_z T_e^{2.5} \frac{\partial T_e}{\partial x} \approx p_e^2 \kappa_o f_z L_z T_e^{0.5} \frac{\partial T_e}{\partial x} \Rightarrow$$

$$\frac{1}{2} dQ_\parallel^2 \approx p_{es}^2 \kappa_o f_z L_z T_e^{0.5} dT_e \Rightarrow \frac{\Delta Q_\parallel}{n_{es} \sqrt{F_z}} \approx \sqrt{2 \bar{\kappa}_o T_{es}^2 \int_0^{T_{es}} L_z(T_e) T_e^{0.5} dT_e}$$

where $\kappa_o \approx \frac{3.1 \times 10^9}{Z_{eff} \ln \Lambda} \left( \frac{erg}{cm\ s\ eV^{3.5}} \right)$ and $F_z(f_z) \equiv \frac{f_z(\%)}{Z_{eff}} = \frac{f_z(\%)}{1 + 0.01 f_z(\%) Z(Z-1)}$

with $\bar{\kappa}_o \equiv \kappa_o Z_{eff}$, $n_e (cm^{-3})$, $T_e$ (eV), $L_z$ ($ergs\ cm^3\ s^{-1}$) and $\Delta Q_\parallel \left( \frac{ergs}{s\ cm^{-2}} \right)$ (2)

In practical units: $\frac{\Delta Q_\parallel \left( \frac{GWatts}{m^2} \right)}{n_{es} (10^{20} m^{-3}) \sqrt{F_z}} \approx 2.5 \times 10^5 \sqrt{T_{es}^2 \int_0^{T_{es}} L_z(T_e) T_e^{0.5} dT_e}$

where the subscript s denotes the value on the separatrix. This integral has been evaluated for six impurities, Be, B, C, Ne, Ar and Kr (Figure 10).

Two candidate machines have been analysed, a 20 MW DIII-D case and a 1.5 GW fusion power ITER case (Table 2). $T_s$ has been determined from parallel heat conduction (equation 3).

$$q_\parallel = -\kappa_o T_e^{2.5} \frac{\partial T_e}{\partial x} = -\frac{2}{7} \kappa_o \frac{\partial T_e^{3.5}}{\partial x} \text{M} T_{es}^{3.5} - T_{div}^{3.5} = \frac{1}{\kappa_o} \int_{div}^{sep} q_\parallel dx \approx \frac{7}{2} L \frac{q_\parallel}{\kappa_o} \text{ with } L \approx 1.4 \pi q_\psi R$$

$$T_{es} \approx 82.2\ eV \left( q_\parallel (GW/m^2)\ R(m)\ q_\psi\ Z_{eff} \right)^{2/7} \text{M where } \kappa_o \approx \frac{3.1 \times 10^9}{Z_{eff} \ln \Lambda} \left( \frac{erg}{cm\ s\ eV^{3.5}} \right)$$ (3)

If the radiation is to be entirely located in the divertor, the upper temperature to be used in the radiation integral (Eq. 2) is the temperature at the X point, for which $L_x \sim 0.3\ L_{tot}$.

Eq. 2 has a different structure than Eq. 1 for edge radiation. $\Delta Q_\parallel$ is proportional to $T_s$ so that larger machines with larger connection lengths have greater radiative capability, but the fusion power in larger machines is also larger. Equations 1 and 2 describe the limits on the effectiveness of exhausting the power by radiation imposed by heat conduction. The volume radiation loss rate scales as $n_e^2 f_z$ and $Q \propto P/R \propto n_e \sqrt{f_z}$, a much weaker dependence. Eq. 1 has $Q \propto [f_z]^{0.5}$, whereas Eq. 2 has $Q \propto [f_z/Z_{eff}]^{0.5}$ which further reduces the radiation efficiency. In addition, the integral in Eq. 2 is weighted by $T^{0.5}$.

Table 2 Typical edge temperatures, connection lengths, and parallel heat fluxes

|  | $P_\alpha$(MW) | $Q_\parallel$(GW/m$^2$) | $A_\perp$(m$^2$) | L (m) | R (m) | $T_s$(eV) | $T_X$(eV) |
|---|---|---|---|---|---|---|---|
| DIII-D 20 MW | 20 | 0.47 | 0.043 | 22 | 1.67 | 120 | 85 |
| ITER 1.5 GW | 240 | 1.5 | 0.16 | 100 | 8.00 | 260 | 185 |

$T_s$ may be higher if the heat flow along the field lines is "flux-limited". The temperature at which "flux limiting" become important can be determined by estimating the flux limited heat flow as

$$q_{\parallel flux-\lim it} = \gamma n_e v_e T_e \approx 6.7 \times 10^{-4} \left( n_e / 10^{20} m^{-3} \right) T_e^{1.5} (eV) \text{ for } \gamma \sim 0.1^{23}.$$ By equating this to the conducted flux in Eq. 3, the condition for the heat flow to be flux limited is approximately

$$T_e \geq 58 eV \left( \frac{n_e}{10^{20} m^{-3}} RqZ_{eff} \right)^{0.5}.$$ For the DIII-D and ITER conditions, $T_{flux\text{-}limit}$ ~ 185 and 400eV respectively, well above the temperatures calculated in Table 2 using heat conduction. Thus the heat flux is not flux limited, especially near the divertor.

The divertor radiation capability of six impurities Be, B, C, Ne, Ar and Kr (Figure 10) indicates that coronal equilibrium rates are inadequate to radiate 20 MW for DIII-D conditions and 240 MW for ITER conditions from the entire scrape-off layer or in the divertor below the X-point(Table 3). An important assumption is that the impurity concentration is uniform throughout the plasma. Strong plasma flows in the divertor plasma would tend to retain and compress the impurities in the divertor chamber. However, the present picture in which the radiating region is upstream of the recycling region would lead to very low plasma flow velocities there so that the thermal force would tend to force impurities toward the main plasma. In addition, plasma flows due to drifts and other effects, and turbulence due to ELM's would tend to lead to mixing of the impurities.

Table 3 Comparison of Be, C, Ne, Ar and Kr divertor coronal equilibrium radiation efficiencies for DIII-D and ITER.

| Element | Be | C | Ne | Ar | Kr |
|---|---|---|---|---|---|
| 0.33 × fatal $f_z$(%) | 4.7 | 2.23 | 0.8 | 0.18 | 0.09 |
| √($f_z$(%)/$Z_{eff}$) | 1.7 | 1.16 | 0.68 | 0.37 | 0.27 |
| $Q_{\parallel DIII-D}$/√($f_z$(%)/$Z_{eff}$) at 120 eV | 0.03 | 0.09 | 0.3 | 0.5 | 0.8 |
| $Q_{\parallel DIII-D}$ for 120 eV and $n_e \approx 10^{20}$ m$^{-3}$ | 0.052 | 0.10 | 0.20 | 0.19 | 0.22 |
| $Q_{\parallel DIII-D}$/√($f_z$(%)/$Z_{eff}$) at 85 eV | 0.018 | 0.05 | 0.2 | 0.25 | 0.3 |
| $Q_{\parallel DIII-D}$ for 85 eV and $n_e \approx 10^{20}$ m$^{-3}$ | 0.031 | 0.058 | 0.14 | 0.094 | 0.081 |
| $Q_{\parallel ITER}$/√($f_z$(%)/$Z_{eff}$) at 260 eV, $n_e \approx 10^{20}$m$^{-3}$ | 0.08 | 0.2 | 0.7 | 2 | 4 |
| $Q_{\parallel ITER}$ (coronal equilibrium) | 0.14 | 0.23 | 0.48 | 0.75 | 1.08 |
| $Q_{\parallel ITER}$/√($f_z$(%)/$Z_{eff}$) at 185 eV, $n_e \approx 10^{20}$m$^{-3}$ | 0.04 | 0.12 | 0.4 | 0.8 | 1.4 |
| $Q_{\parallel ITER}$ (coronal equilibrium) | 0.07 | 0.14 | 0.27 | 0.3 | 0.38 |

The radiation level is proportional to $T_s \sqrt{\int_0^{T_s} L_z T^{0.5} dT}$ so that enhancement of the emissivity L would increase the radiative losses. L can be enhanced by charge exchange recombination and impurity recycling by altering the ionisation balance toward lower charge states which radiate more strongly[16,24-26]. The charge exchange recombination rate coefficient, $<\sigma v>_{CX} n_o n_i = <\sigma v>_{CX} n_e n_i (n_o/n_e)$, scales like electron-ion recombination and ionisation with an extra factor of $n_o/n_e$, the fraction of neutral atoms to the electron density, so that the charge state distribution, and therefore L(T), can be parametrized by $n_o/n_e$. Similarly, because the ionisation and recombination rates have the form $<\sigma v> n_e n_z^{i+}$, $n_e$ can be factored out to produce an equation for the ionisation balance of the form $\partial n_z^{i+}/\partial(n_e t) = F(<\sigma v> n_z^{i+})$, so that the effect of rapid recycling can be characterised by $n_e \tau_{recycle}$. The magnitude of effect can be estimated by integrating this equation with the initial conditions: $n_z^{0+}(n_e t=0) = n_{tot}$, $n_z^{i+} = 0$ for i=1,Z, from t=0 to t=$\tau_{recycle}$, and parameterizing the radiation equilibrium and emissivity as a function of $T_e$ for different values of $n_e \tau_{recycle}$ [27]. Using the same formalism as with coronal equilibrium, one can evaluate the $n_o/n_e$ and $n_e \tau_{recycle}$ required to enhance the radiation rate to radiate the heating power. Approximately $n_o/n_e \approx 10^{-2}$ to $10^{-3}$ and $n_e \tau_{recycle} \approx 5 \times 10^{15}$ to $5 \times 10^{16}$ m$^{-3}$ s is required for impurities (with $n_s \approx 10^{20}$ m$^{-3}$) to the power in the scrape-off layer for both the DIII-D 20 MW case and the ITER 240 MW case (Figure 11—12, Table 4). Neon appears to be the optimum impurity in that it has the least demanding requirements, i.e. the lowest $n_o/n_e$ and the largest $n_e \tau_{recycle}$.

Table 4 Comparison of the requirements for enhancement of Be, C, Ne, and Ar divertor radiation efficiencies for DIII-D and ITER.

| Element | Be | C | Ne | Ar |
|---|---|---|---|---|
| $n_o/n_e$ required to radiate 20 MW (DIII-D) | $5 \times 10^{-2}$ | $10^{-2}$ | $10^{-2}$ | $5 \times 10^{-2}$ |
| $n_e \tau_{recycle}$ (s m$^{-3}$) required to radiate 20 MW (DIII-D) | $10^{16}$ | $10^{16}$ | $3 \times 10^{16}$ | $5 \times 10^{15}$ |
| $n_o/n_e$ required to radiate 240 MW (ITER) | $7 \times 10^{-3}$ | $8 \times 10^{-3}$ | $10^{-3}$ | $4 \times 10^{-2}$ |
| $n_e \tau_{recycle}$ (s m$^{-3}$) required to radiate 240 MW (ITER) | $10^{16}$ | $10^{16}$ | $4 \times 10^{16}$ | $6 \times 10^{15}$ |

## 6. Summary

Transferring the energy from the plasma to the plasma facing components by atomic processes is a promising approach for reducing the peak loads for the next generation of fusion experiments such as ITER. To better understand the results of the complex simulations and experiments, we have used simple models to examine the potential role of central radiation losses through bremsstrahlung, radiation losses from the plasma edge, charge exchange hydrogen radiation and losses from the scrape-off layer and divertor plasma and impurity radiation losses from the divertor plasma. The simple analysis gives results which are consistent with the main featurea of the complex models. The conclusion of the simple models is that each process can contribute to the solution of the problem, with major roles being played by Bremsstrahlung from the plasma core and impurity radiation from the plasma edge and divertor plasma with smaller contributions due to charge exchange and hydrogen radiation losses. To achieve the required levels of impurity radiation in the divertor will likely require enhancements of the impurity radiation due to charge exchange recombination with $n_o/n_e \approx 10^{-2}$ to $10^{-3}$ and impurity recycling levels of the order of $n_e\tau_{recycle} \approx 5 \times 10^{15}$ to $5 \times 10^{16}$ m$^{-3}$ s. Neon appears to be the optimum impurity in that it requires the lowest $n_o/n_e$ and the largest $n_e\tau_{recycle}$.

## Acknowledgements

The authors gratefully acknowledge discussions with members of the ITER Joint Central Team, including G. Janeschitz, F. Perkins and M. Rosenbluth and other members of the divertor community, including S. Allen, S. Cohen, D. Hill, M. Kaufmann, K. Lackner and R. Stambaugh.

**Figures:**

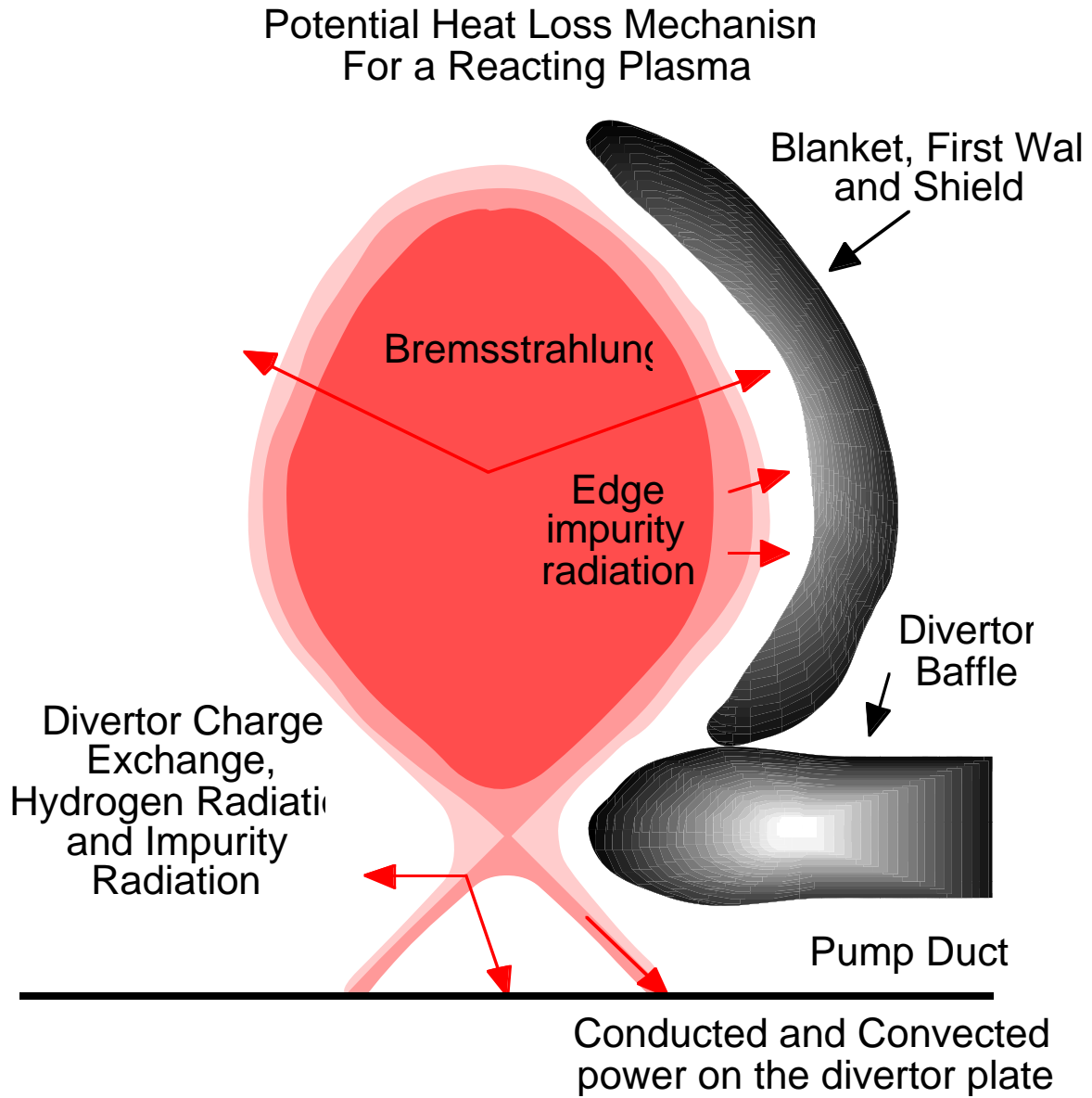

Figure 1. Schematic illustration of energy loss mechanisms in tokamaks due to atomic processes.

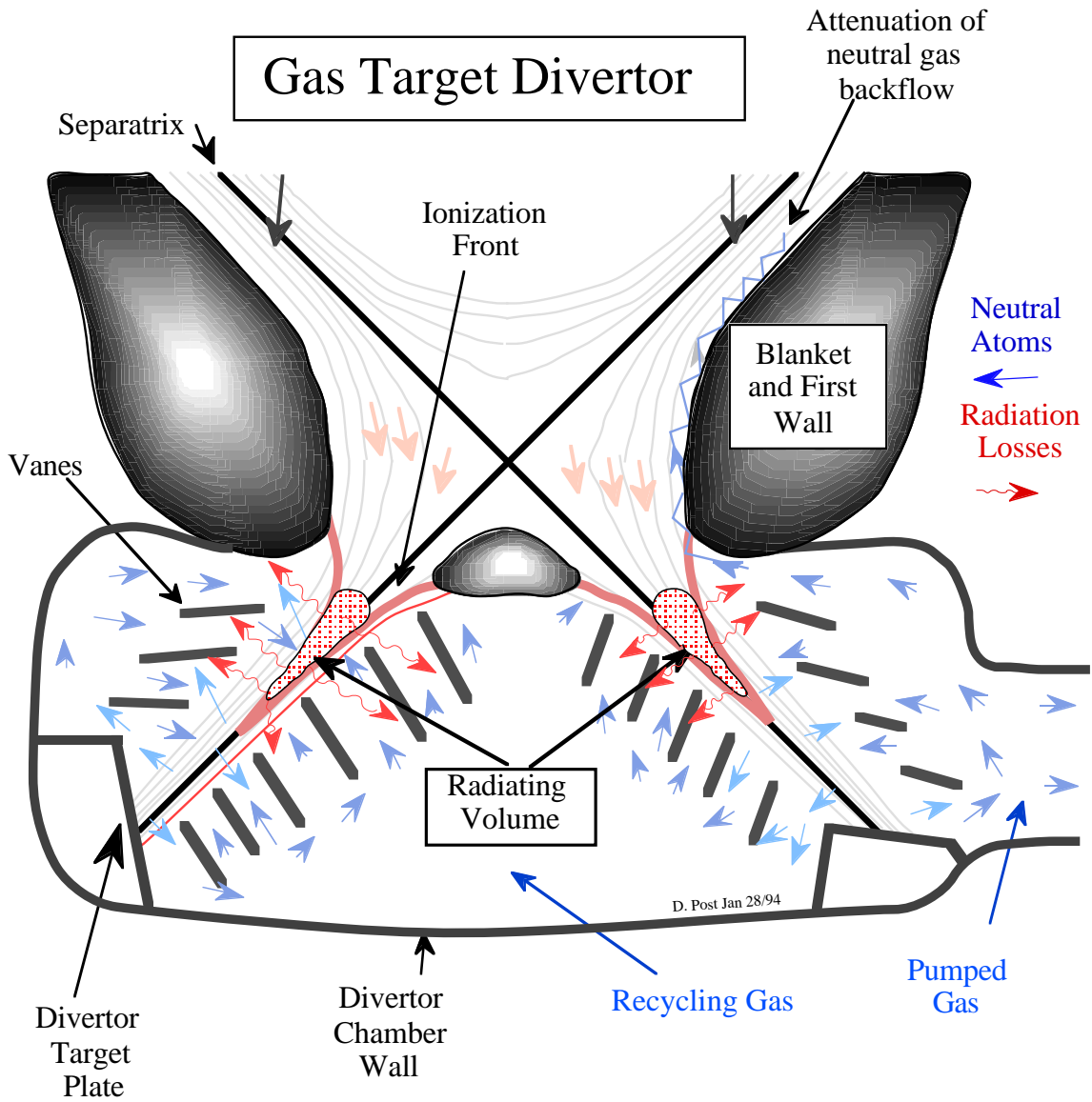

Figure 2. Schmatic Illustration of Dynamic Gas Target Divertor.

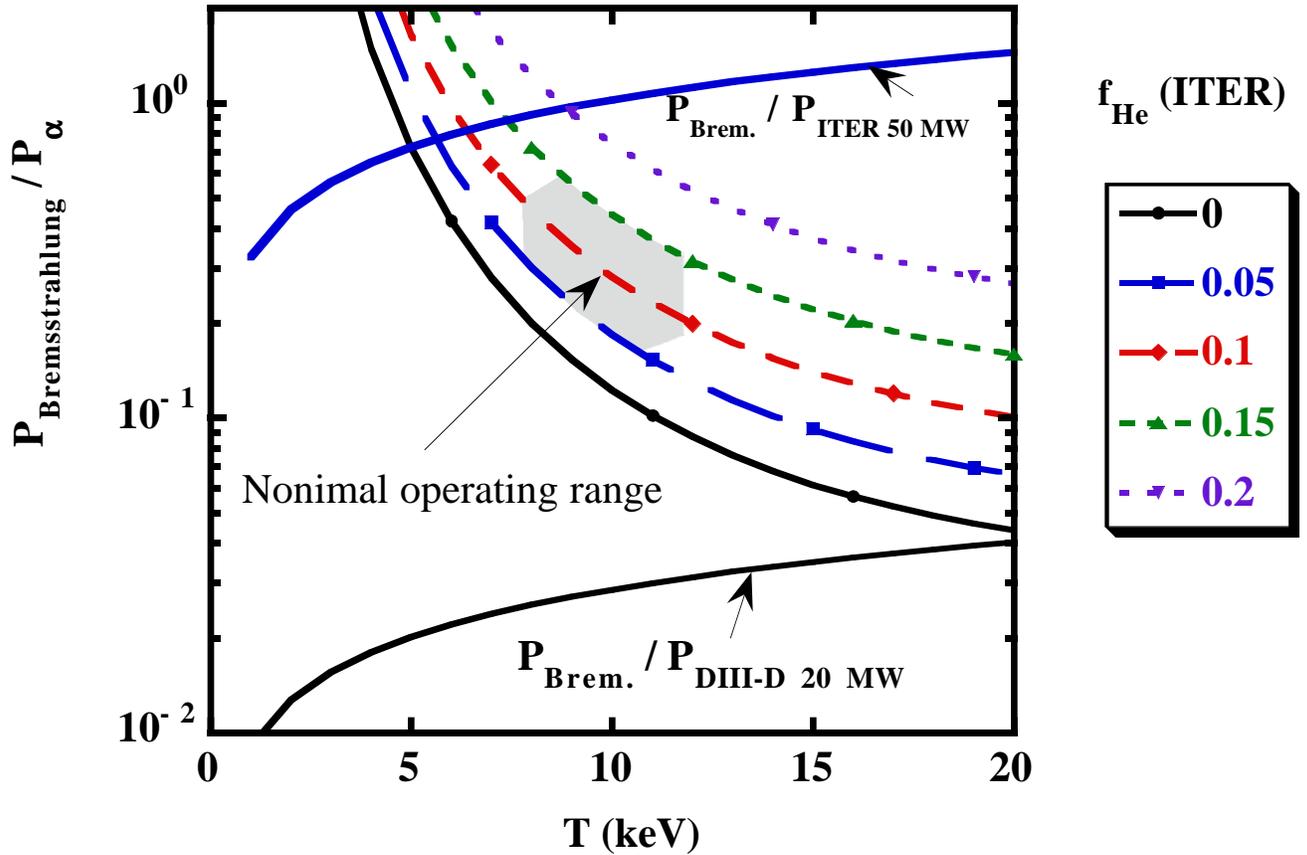

Figure 3. $P_{Bremsstrahlung} / P_\alpha$ for ITER as a function of Temperature for various values of $f_{He}$ with $f_{Be} = 0.1\, f_{He}$. The nominal operating range for ITER ($8\text{ keV} \leq \langle T \rangle \leq 12\text{ keV}$, and $0.05 \leq f_{He} \leq 0.15$ is shaded. Value of $P_{Bremsstrahlung} / P_{heat}$ are shown for 20 MW of heating in DIII-D ($f_C \approx 0.01$) and 50 MW of heating in ITER ($f_{Be} \approx 0.02$).

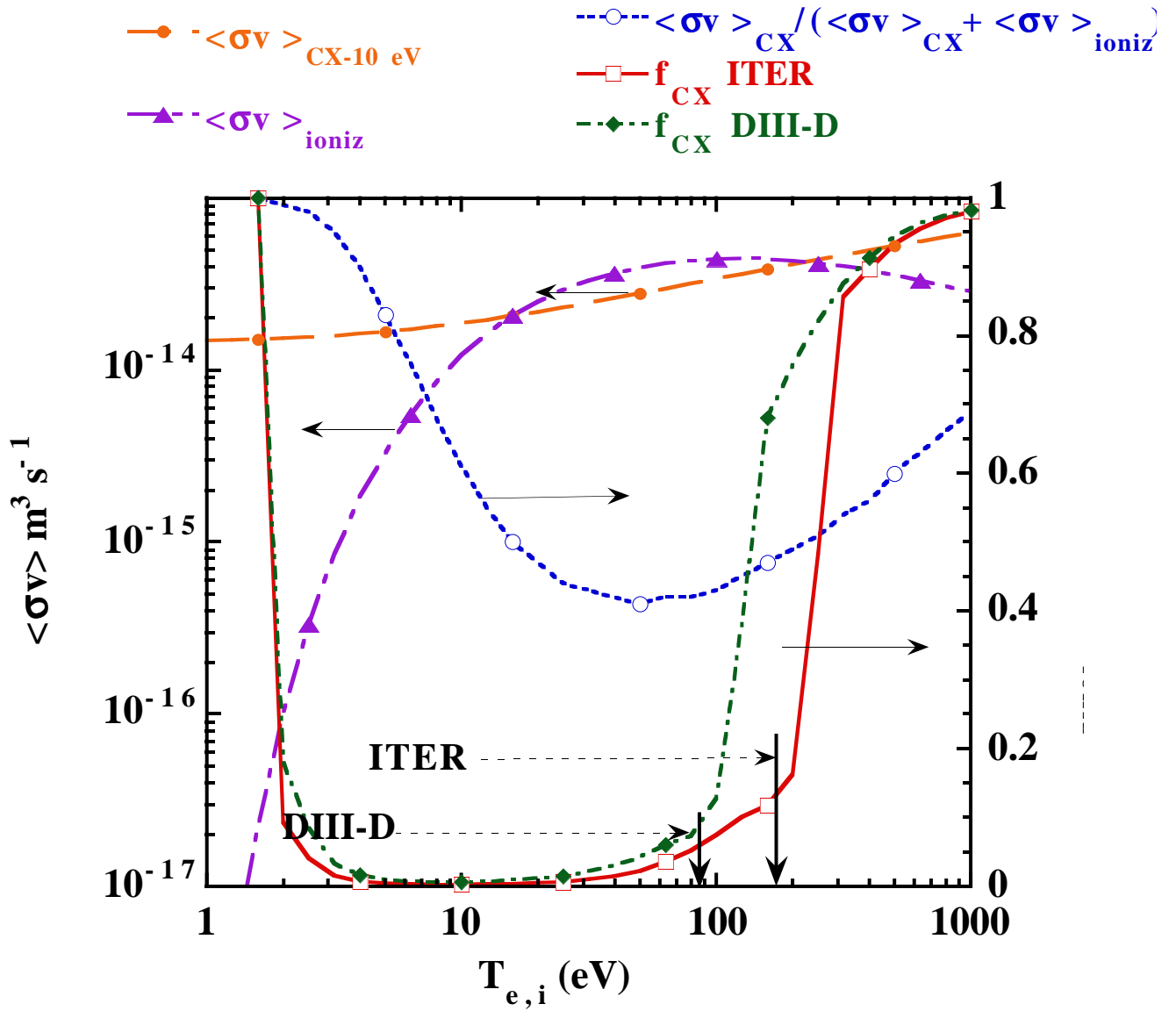

Figure 4. The fraction of charge exchange losses $f_{CX}$ for DIII-D and ITER conditions, $\langle \sigma v \rangle$ for ionization and charge-exchange for $n_e = 10^{20}$ m$^{-3}$ and $\langle \sigma v \rangle_{CX}/(\langle \sigma v \rangle_{CX} + \langle \sigma v \rangle_{ionization})$.

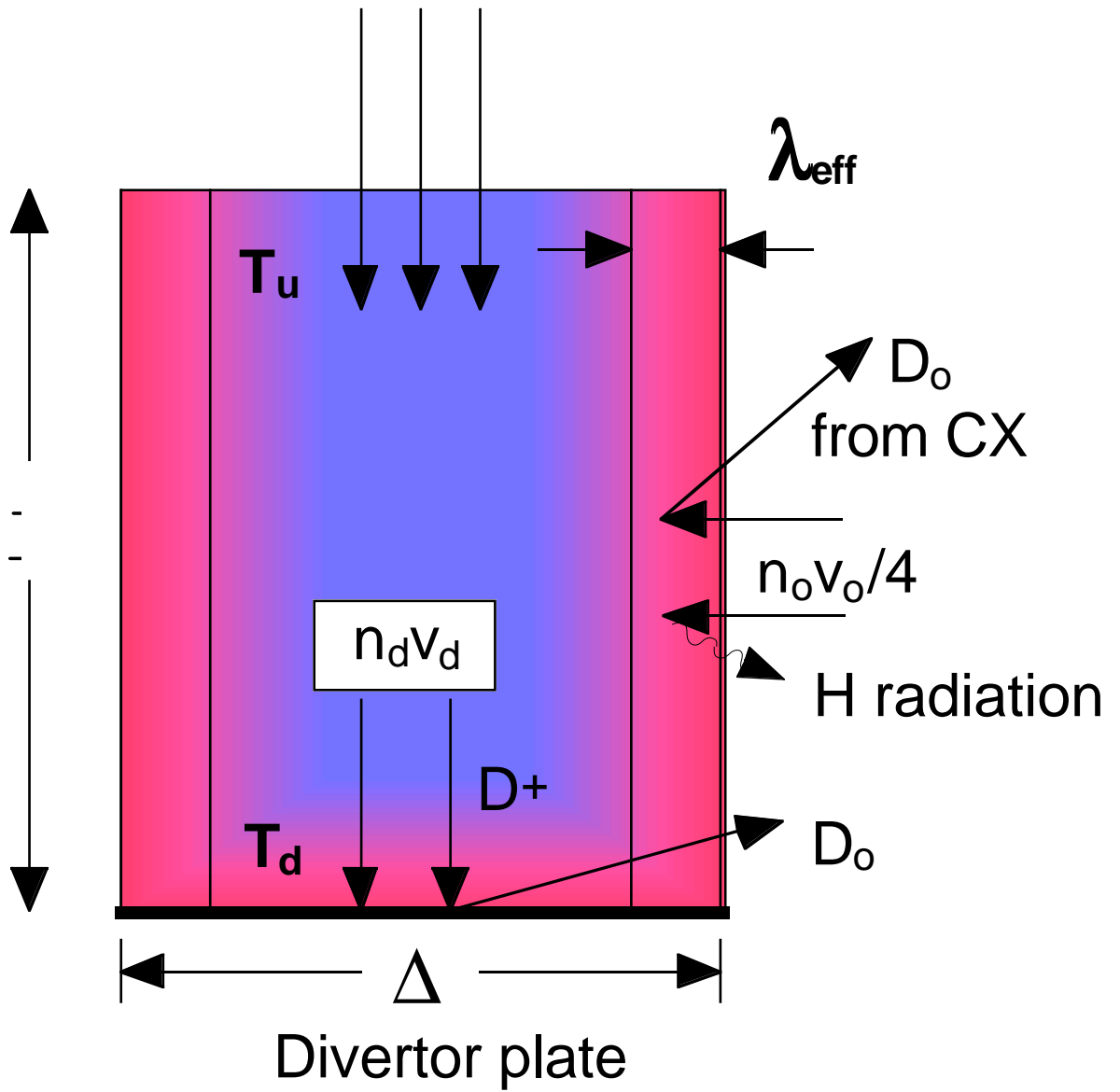

Figure 5. Schematic illustration of the power and particle balance in a divertor illustrating the roles of charge exchange and hydrogen radiation losses from the upstream divertor plasma (away from the divertor plate).

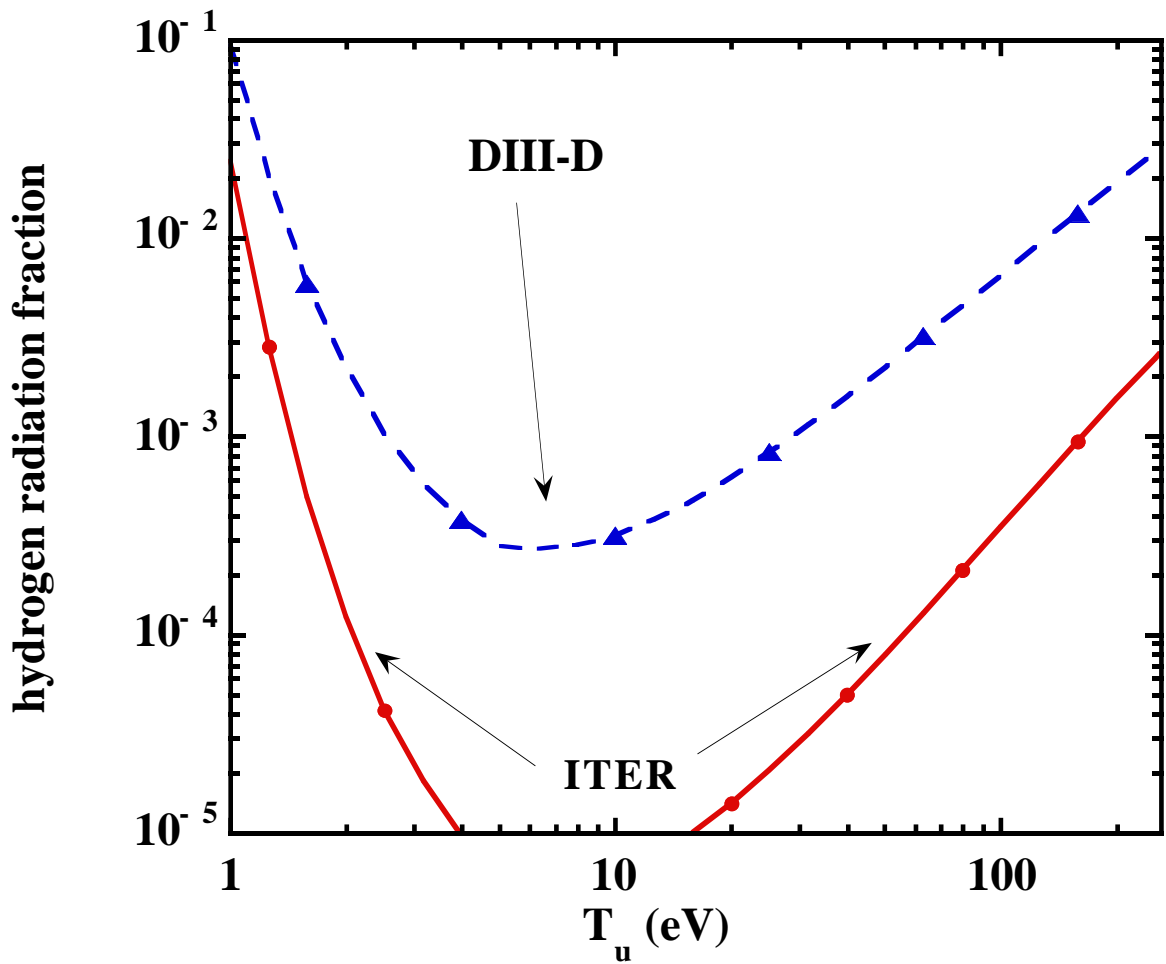

Figure 6. Ratio of the hydrogen radiated power losses from the upstream divertor plasma to the power incident on the divertor plate for ITER and DIII-D conditions.

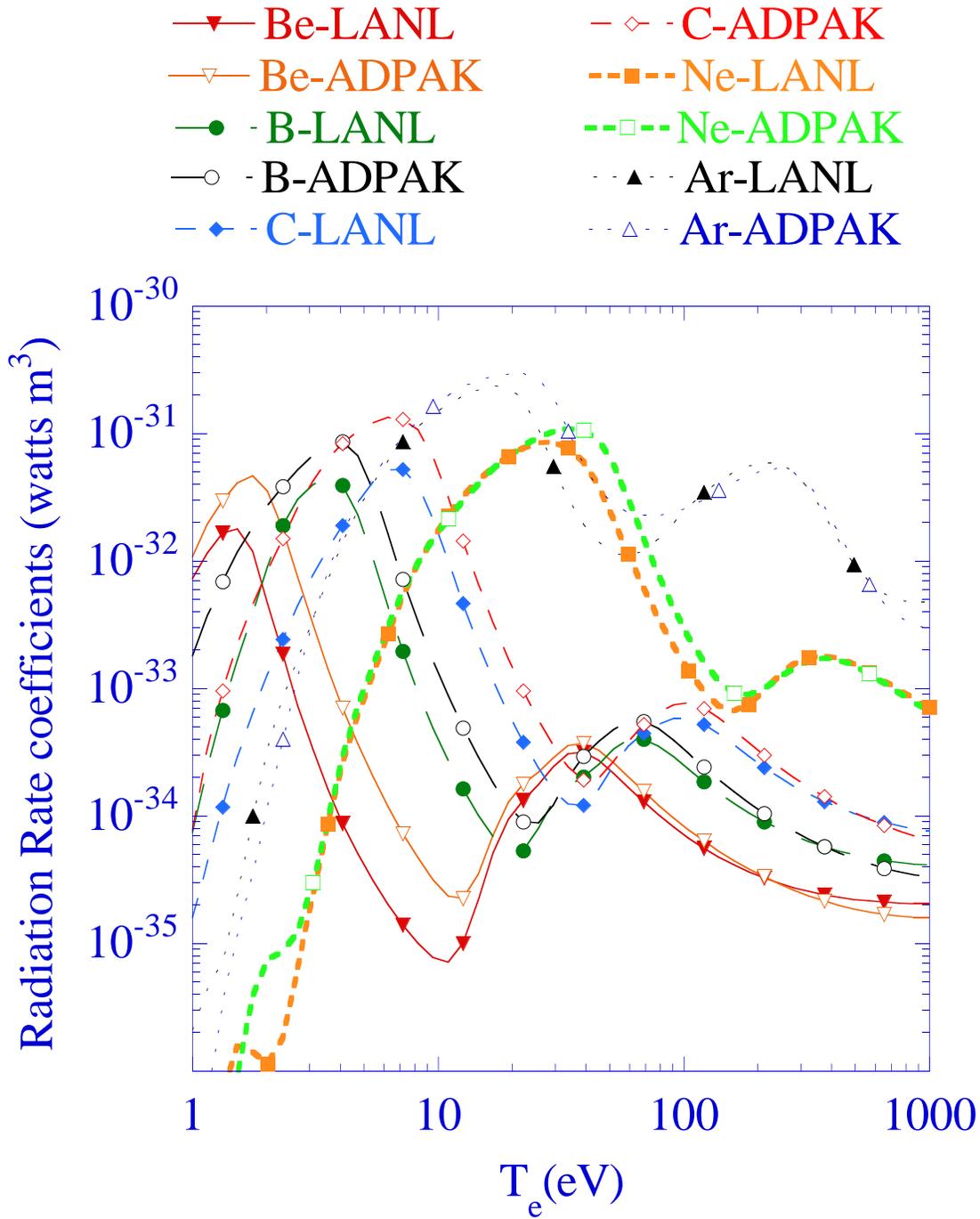

Figure 7. Comparison of Radiation rate coefficients for Be, B, C, Ne and Ar for ADPAK and LANL calculations.

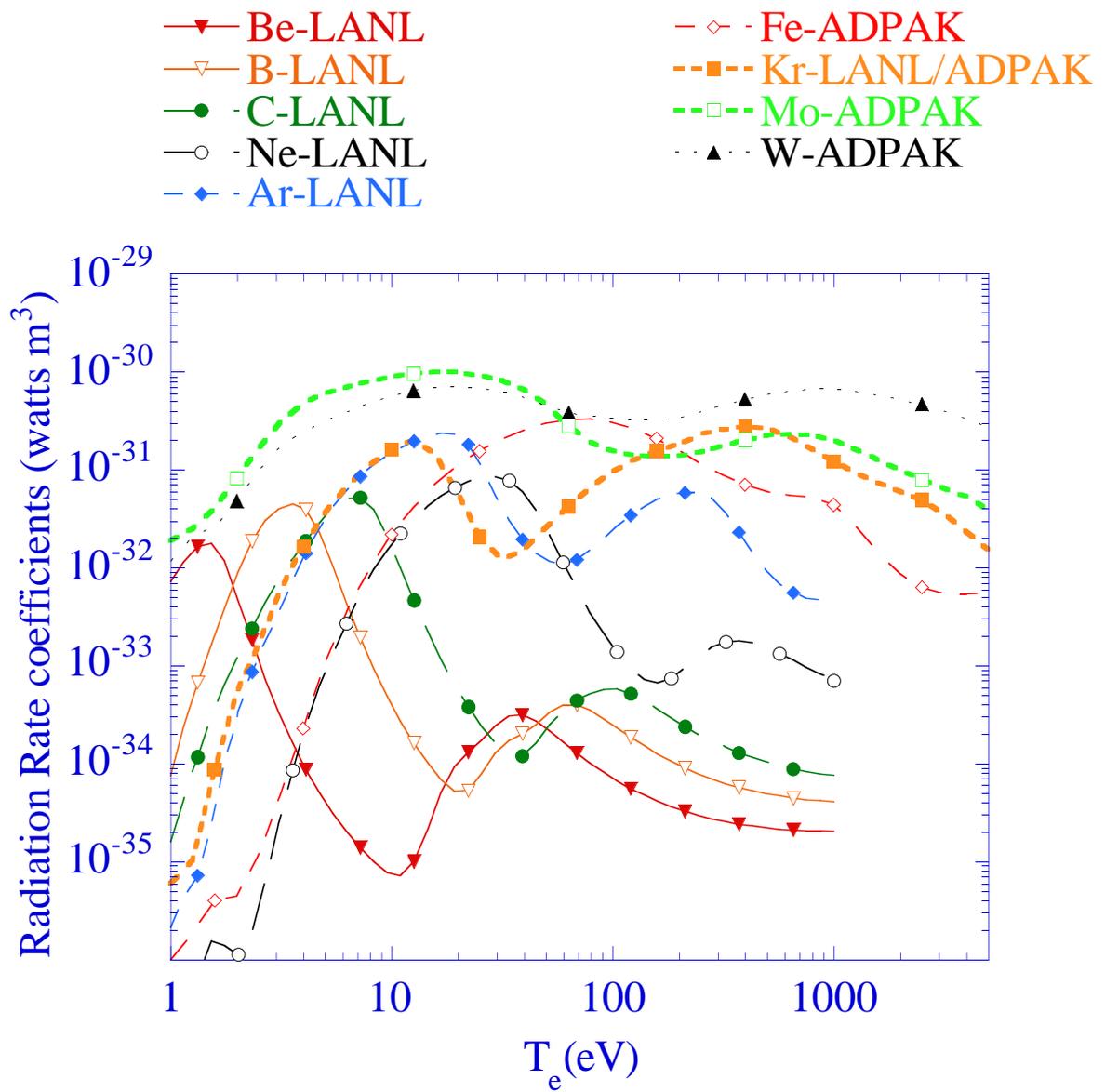

Figure 8. Radiation loss rate coefficients for Be, B, C, Ne, Ar, Fe, Kr, Mo and W.

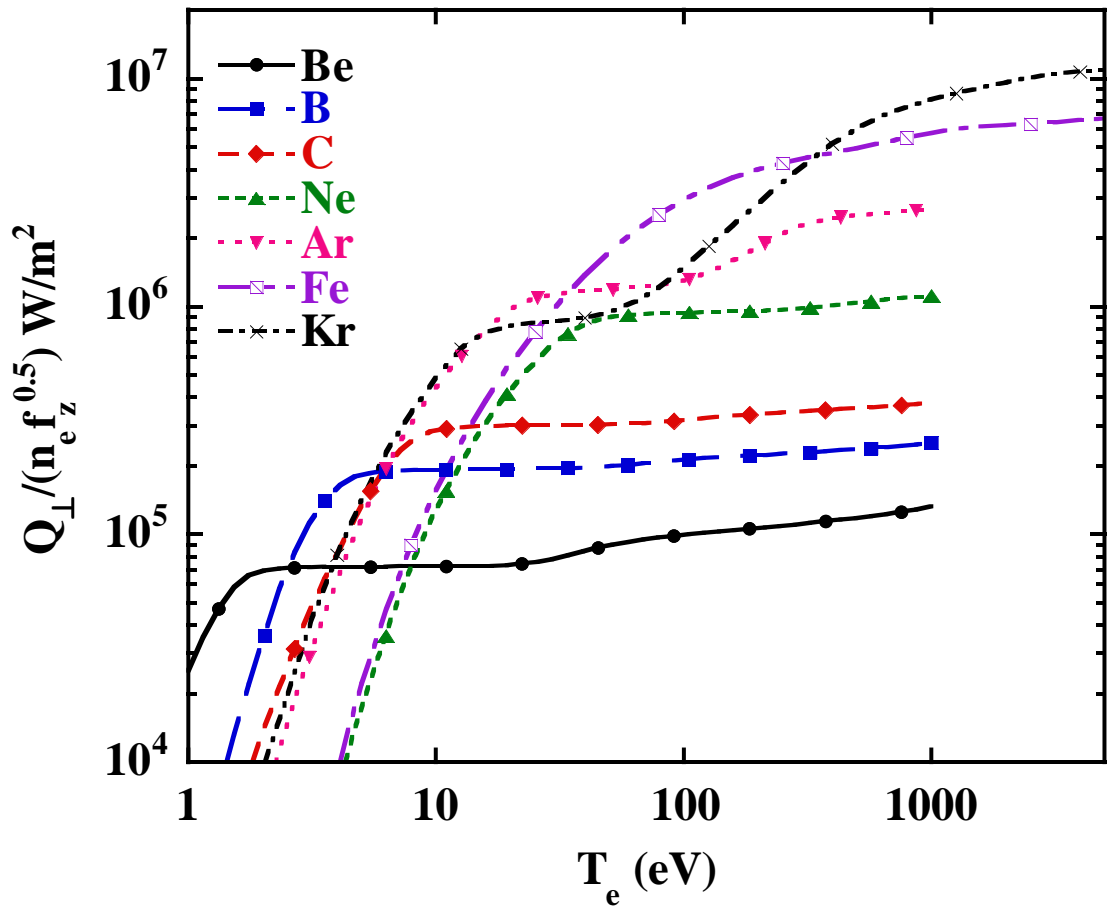

Figure 9. Normalized $Q_\perp$ for the main plasma edge as a function of $T_e$ for Be, B, C, Ne, Ar, Fe and Kr.

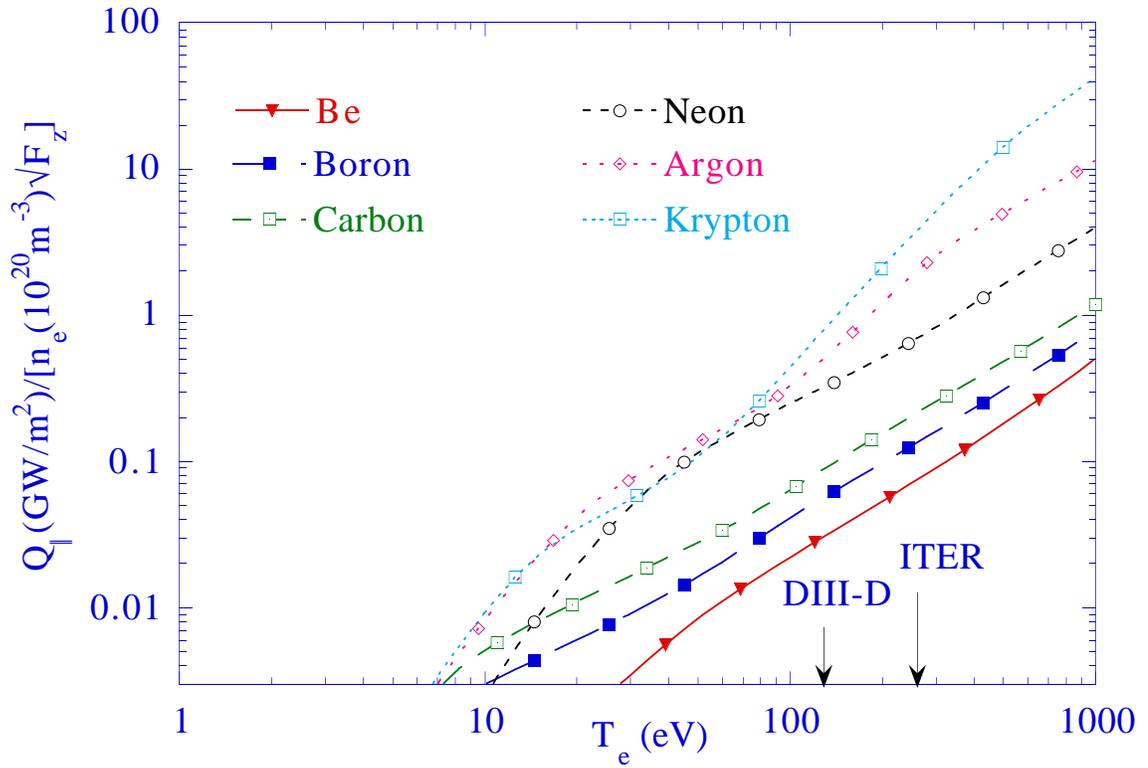

Figure 10. Normalized $Q_\parallel$ for the divertor plasma and Scrape-Off-Layer for Be, B, C, Ne, Ar and Kr as a function of the temperature at the mid-plane separatrix.

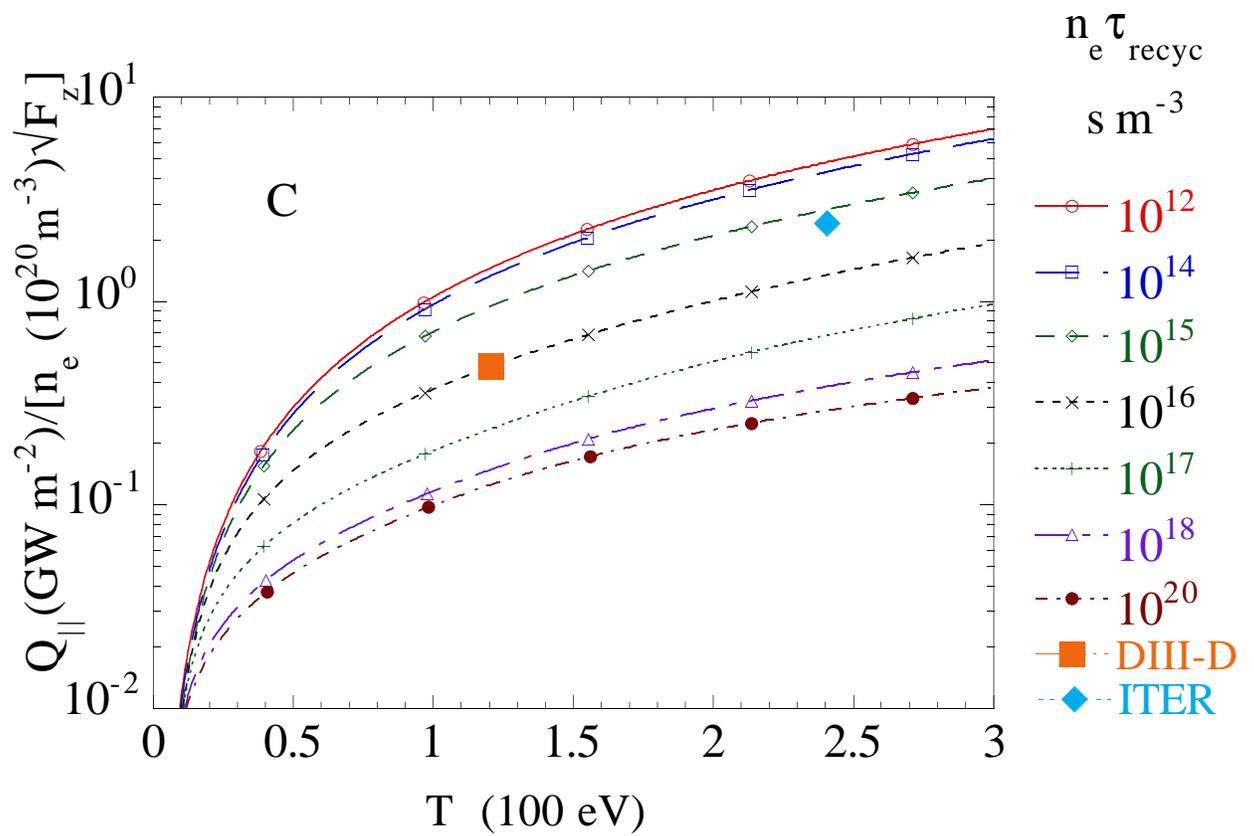

Figure 11. Normalized $Q_\parallel$ for the divertor plasma and Scrape-Off-Layer for C as a function of the temperature at the mid-plane separatrix and recycling level ($n_e\tau_{recycle}$)

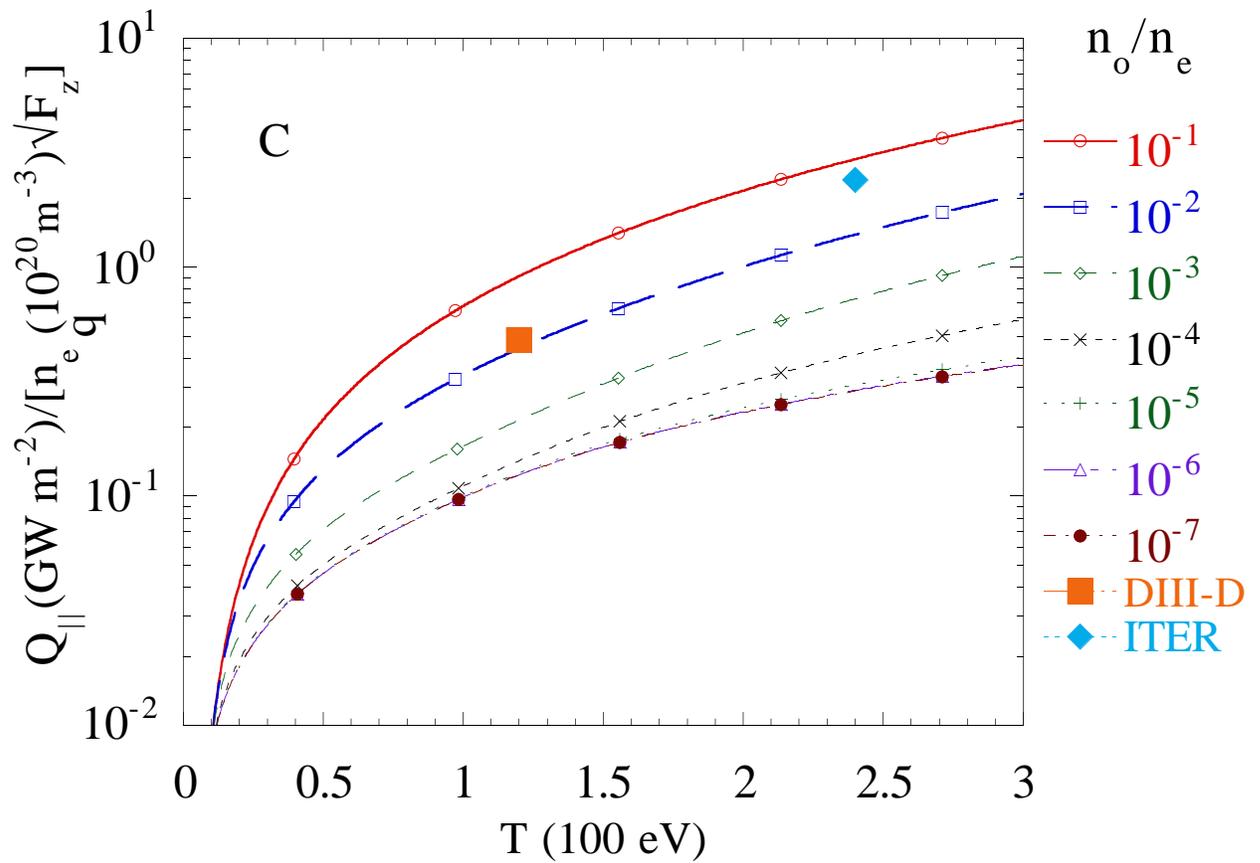

Figure 12. Normalized $Q_{\parallel}$ for the divertor plasma and Scrape-Off-Layer for C as a function of the temperature at the mid-plane separatrix and neutral fraction $n_o/n_e$.